\documentclass[preprint]{revtex4}

\usepackage{graphicx}

\newcommand{\beq}{\begin{equation}}
\newcommand{\enq}{\end{equation}}
\newcommand{\bea}{\begin{eqnarray}}
\newcommand{\ena}{\end{eqnarray}}
\newcommand{\psid}{\psi^{\dagger}}
\newcommand{\rr}{{\bf r}}
\newcommand{\RR}{{\bf R}}
\newcommand{\aos}{a_{\mathrm{osc}}}

\begin{document}
\title{Center of mass rotation and vortices in an attractive Bose gas}
\author{A. Collin$^1$, E. Lundh$^{2,1}$, K.-A. Suominen$^{3}$}
\affiliation{$^1$Helsinki Institute of Physics, PL 64, FIN-00014
Helsingin yliopisto, Finland\\
$^2$Condensed matter theory group,
Royal Institute of Technology, SE-10691 Stockholm, Sweden\\
$^3$Department of Physics, University of Turku, FIN-20014
Turun yliopisto, Finland}
\date{\today}

\begin{abstract}
The rotational properties of an attractively interacting Bose gas are
studied using analytical and numerical methods. We study perturbatively
the ground state phase space for weak interactions, and find that in an
anharmonic trap the rotational ground states are vortex or center of mass
rotational states; the crossover line separating these two phases is
calculated. We further show that the Gross-Pitaevskii equation is a valid
description of such a gas in the rotating frame and calculate numerically
the phase space structure using this equation. It is found that the
transition between vortex and center of mass rotation is gradual;
furthermore the perturbative approach is valid only in an exceedingly
small portion of phase space. We also present an intuitive picture of the
physics involved in terms of correlated successive measurements for the
center of mass state.
\end{abstract}

\maketitle

\section{Introduction}
The rotational properties of atomic Bose condensed gases have been a 
matter of intense scientific interest recently. The main role
in these studies is played by vortices, which are the manifestations of 
quantized circulation in the quantum gas in question. 
However, vortices are by no means the only possible rotational states.

By rotating a condensate in a harmonic trap with a certain frequency $\Omega$
a vortex can be created; when rotating the condensate a bit faster 
a second vortex appears, and so on~\cite{Madison2000}.
Lattices with a large number of vortices have been experimentally
observed in many groups~\cite{Abo2000,Engels2003}.
There is an upper limit, however, for the rotation frequency in a harmonic
trap. This limit is the trap frequency $\omega$ of the
confining trap, above which the center of mass of the condensate is
destabilized~\cite{Rose2002}. Interesting physics has been done by
the Boulder group working only slightly below this limit~\cite{Engels2003}.
Having prepared a condensate with a vortex lattice, atoms with low
angular momenta are then removed.
Reaching the rotation rate $0.95\omega$
a metastable giant vortex with a phase singularity on the order of
50 is observed.

On the other hand, the trap limit
can be crossed if the trap is not purely harmonic but the potential
has a quartic term also \cite{Fetter2001}. 
The steeper potential makes fast rotation
possible and thus gives a tool to test theoretical predictions including
multiply quantized vortices \cite{Lundh2002}, strongly correlated vortex
liquids or Quantum Hall like states \cite{Cooper2001,Ho2001}.
The first experimental studies have already
been done at ENS~\cite{Bretin2003}.

The rotation of a condensate in an anharmonic trap is especially interesting
when the effective interactions of the atoms are {\it attractive}.
It was shown by Wilkin {\it et al.} that for weak attractive
interactions and a given angular momentum, in the lowest energy
state the motion is carried by the center of mass (c.~m.)~\cite{Wilkin1998}.
For this state the single-particle reduced density matrix has more than 
one macroscopic eigenvalue in the laboratory reference of frame.
This might indicate that the c.m. rotational state is a fragmented condensate,
but actually, the state is Bose condensed if it is
viewed in the c.~m.\ frame~\cite{Pethick2000}. Unfortunately,
in a harmonic-oscillator trap
this state is never thermodynamically stable because the critical
frequency for the excitation is equal to the frequency where the
gas is destabilized. An anharmonic term in
the trapping potential removes the instability and makes the state of c.~m.\
rotation, as well as vortex states, attainable~\cite{Us2003}.

In this paper we shall study the rotational states of an attractive
zero temperature Bose gas confined in an anharmonic trap. We begin by
discussing the nature of the relevant states of motion and perturbation theory
in Sec.~\ref{sec:perturbation}. In Sec.~\ref{sec:equations} we discuss the
validity of different equations of motion. The problem is approached
within the ordinary Gross-Pitaevskii (GP) mean field theory
in Sec.~\ref{sec:numerics}. Section \ref{sec:correlations} deals with the
correlation properties of the c.~m.\ rotational state at the limit of small
anharmonicity. Concluding remarks are made in Sec.~\ref{sec:conclusions}.

\section{Hamiltonian and perturbation theory}
\label{sec:perturbation}

The Hamiltonian for $N$ atoms with binary {\it s}-wave interactions is
\beq\label{hamiltonian}
   H = -\frac{\hbar^2}{2m}\sum_{i=1}^N\frac{d^2}{d\rr_i^2}
   +\sum_{i=1}^N V(\rr_i)
   + \frac{U_0}{2}\sum_{i\neq j} \delta(\rr_i-\rr_j)
\enq
where the strength of the contact interaction
$U_0=4\pi\hbar^2a/m$ with the scattering length $a$ which is here
assumed negative. The external field $V$ consists of a
cylindrically symmetric harmonic potential with a small quartic
addition in the radial direction:
\beq\label{anharmpotential}
   V(r,\theta,z)=\frac12m\left[\omega^2 \left(r^2 +
   \lambda \frac{r^4}{\aos^2}\right)
   +\omega_z^2 z^2\right].
\enq
Here the dimensionless parameter $\lambda$ describes the
strength of the anharmonic term and the oscillator length $\aos$
is defined as $\aos=(\hbar/m\omega)^{1/2}$.

In Refs.\ \cite{Wilkin1998,Mottelson1999}, the harmonically
trapped gas ($\lambda=0$) was studied by treating the interaction
energy perturbatively, in which case the many-body eigenstates are
products of nodeless
harmonic-oscillator single-particle eigenstates. Here
we shall generalize this approach by including also the anharmonic
potential~\footnote{Since the first submission for publication of the 
present manuscript, the analysis in the present chapter has been 
extended by Kavoulakis {\it et al.} \cite{Kavo2004}.}. 
To this end, we first pass to dimensionless units,
where lengths are scaled by $\aos$ and
energies by $\hbar\omega$. We obtain
\beq\label{dimlesshamiltonian}
H = -\frac12 \sum_{i=1}^N\frac{d^2}{d\rr_i^2}
+\frac12 \sum_{i=1}^N \left(r_i^2 + \lambda r_i^4\right)
+ \frac{4\pi g}{2}\sum_{i\neq j} \delta(\rr_i-\rr_j),
\enq
where we have defined an effective 2D interaction $g$ by
integrating out the $z$-component wave function to obtain
$g=\sqrt{\omega_z/(2\pi\omega)} a/\aos$ (this can be done because
the radial-axial
decoupling is exact in perturbation theory). In the attractive case,
$g$ is negative. We shall retain the
symbols $\rr$, $H$ etc.\ for the dimensionless quantities in this
section, because no confusion can arise.

A general form for a many-body state in two dimensions is
\beq
\Psi(\rr_1,\ldots \rr_N) =
\sum_{m_1,\ldots,m_N}c_{m_1,\ldots,m_N} \prod_{j=1}^{N}
\phi_{m_j,0}(\rr_j),
\enq
where $\phi_{mn}(\rr)$ is the harmonic-oscillator single-particle
eigenstate in two dimensions with angular momentum $m$ and radial
quantum number $n$; only states with $n=0$ participate in
perturbation theory. In the unperturbed case ($\lambda=a=0$), for
a given total angular momentum $L$ all states which fulfill
$m_1+\ldots+m_N=L$ with all $m_j\geq 0$ (and have no radial nodes,
i.~e.\ $n=0$) are degenerate. Thus we have, in principle, to perform
degenerate perturbation theory on a vast subspace of Hilbert space
(the number of basis states is $(N+L-1)!/[N!(L-1)!]$). However,
from physical arguments and known exact results, we obtain
guidance to what the relevant candidate states should be.

In the purely harmonic case with finite attractive interactions,
the ground-state many-body wave function
with angular momentum $L$ and particle number $N$ was found to
be~\cite{Wilkin1998}
\beq\label{wilkinwf}
\Psi^C_L(z_1,\ldots,z_N) = \sqrt{\frac{N^L}{\pi^N L!}}
\left(\sum_{j=1}^N \frac{z_j}{N}\right)^L \exp\left({-\sum_{j=1}^N |z_j|^2/2}\right),
\enq
where $z_j=x_j+iy_j$ are the particle coordinates.
The quantity within parentheses is the c.~m.\ coordinate; therefore,
this wave function describes rotation of the c.~m.\, as already
mentioned in the introductory paragraph. We consider this state as 
the {\it ideal}~c.~m.\ rotational state.

When instead the anharmonicity $\lambda$ is nonzero and
the interactions vanish, it was shown in Ref.\ \cite{Lundh2002}
that the ground state is the vortex state
\beq\label{vortexwf}
\Psi^v_m(z_1,\ldots,z_N) = \prod_{j=1}^N \frac{1}{\sqrt{\pi m!}}
z_j^m e^{-|z_j|^2/2},
\enq
which is just a Bose-Einstein condensate containing an $m$-fold
quantized vortex, thus having angular momentum $Nm$.
Vortex-array states do not come into question because they have a lower
density on the average, and are therefore not favored by the attractive
interaction.

It should be noted that Eqs. (\ref{wilkinwf}) and 
(\ref{vortexwf}) are the extreme cases where either
the interaction or anharmonicity are absent. In a more general context we
make the difference between these states in terms of cylindrical symmetry.
The vortex state is cylindrically symmetric and the ideal c.~m.\ rotational
state corresponds clearly to the nonsymmetric case. Between these two
states, however, there is a region of nonsymmetric cases, where the c.~m.\ 
is in rotation although these states do not correspond to the ideal c.~m.\
rotational case of Eq. (\ref{wilkinwf}). The leading instability
of the vortex state towards a nonsymmetric state was examined in Ref.
\cite{Kavo2004}. We include these cylindrically nonsymmetric states into 
our general definition of c.~m.\ rotational states. 
Other interpretations are possible, though, see \cite{Kavo2004}.

There now remains to compare the energies of the two
states (\ref{wilkinwf}) and (\ref{vortexwf})~\footnote{Reference
\cite{Mottelson1999} discusses other types of angular-momentum carrying
states; these are, however, easily seen to always have higher energy than
(\ref{wilkinwf}) and (\ref{vortexwf}).}.
The interaction energies of the two states were calculated in
Ref.\ \cite{Mottelson1999} and amount to
\bea
\langle \Psi^C_L |H_{\rm int}|\Psi^C_L\rangle &=& gN(N-1),\nonumber\\
\langle \Psi^v_m |H_{\rm int}|\Psi^v_m\rangle &=&
gN(N-1)\frac{(2m)!}{(m!)^2 2^{2m}}\\
&=&gN(N-1)\frac{(2m-1)!!}{(2m)!!}\nonumber.
\ena
The absolute magnitude of the interaction energy in the vortex state
is smaller than that of
the ideal c.~m.\ state, and therefore the latter is favored by the
attractive interactions (remember that $g<0$).
The quartic energies, on the other hand, are proportional to
\bea
\langle \Psi^C_L |\sum r_j^4|\Psi^C_L\rangle &=& N\left(2 + 4\frac{L}{N}
+\frac{L^2}{N^2} -\frac{1}{N}\frac{L}{N}\right),\nonumber\\
\langle \Psi^v_m |\sum r_j^4|\Psi^v_m\rangle &=& N(m+1)(m+2),
\ena
so that for equal angular momentum, the quartic energy of the vortex
state is the lowest.

In the vortex state the angular momentum per particle $m$ can only
take on integer values. In contrast, in the ideal c.~m.\ state, $q=L/N$ is
only quantized in quanta of $1/N$, so when we pass to the infinite
limit $q$ is a continuous variable. At fixed $L$, it is easy to
determine the boundary between the rotational phases: the ideal
c.~m.\ state is energetically favorable when
\beq\label{fixedl}
\frac{\lambda}{2}m < |g|N\left(1 - \frac{(2m-1)!!}{(2m)!!}
\right).
\enq
We have discarded all terms ${\mathcal O}(N^{-1})$. For $m=q=1$
the condition for c.~m.\ rotation is $\lambda < |g|N$ and for faster
rotation, the regime where the ideal c.~m.\ state is favorable is smaller.
Since the right-hand side of Eq.\ (\ref{fixedl}) increases slower
than the left-hand side as $m$ increases, we conclude that for
any given pair of parameters $(\lambda,gN)$, there exists a
smallest angular momentum $L$ above which a vortex state is
favorable.

We now change variables and work at fixed angular velocity $\Omega$
rather than fixed
$L$. In the ideal c.~m.\ state we can simply differentiate the energy:
\beq\label{pertangmo}
\Omega = \frac{dE}{dL} = 1 + \lambda(2+q) \Rightarrow q =
\frac{\Omega-1}{\lambda} - 2.
\enq
The critical frequency for the excitation of rotational motion is
thus $\Omega_C^{cm}=1+2\lambda$, and above this threshold we substitute
the above expression for $q$ in order to obtain the energy as a function
of $\Omega$:
\beq\label{ecomega}
\frac{1}{N}[E_{C}(\Omega,\lambda,g) - \Omega L]  = -1 + 2\Omega
-\frac{(\Omega-1)^2}{2\lambda} - \lambda -|g|N.
\enq
The energy of the vortex state as a function of $m$ is
\beq\label{evomega}
\frac{1}{N}E_v(m) =  1 + m + \frac12\lambda(m+1)(m+2) -
|g|N \frac{(2m-1)!!}{(2m)!!}.
\enq
Comparing Eqs.\ (\ref{ecomega}) and (\ref{evomega}), the
critical angular velocities for the successive vortex states are
found to be
\bea
\Omega^v_{C1} & = & 1 + 2\lambda +\frac12 |g|N,\nonumber\\
\Omega^v_{C2} & = & 1 + 3\lambda +\frac18 |g|N,\nonumber\\
\Omega^v_{C,m}& = & 1 + (m+1)\lambda +
|g|N\frac{(2m-3)!!}{(2m)!!}.
\ena
For given $gN$, $\lambda$ and $\Omega$ it can now be determined which of the
states $\Psi^C$ and $\Psi^v$ has the lower energy.
When $\Omega^v_{C1} < \Omega < \Omega^v_{C2}$, we find that the ideal 
c.~m.\ state is favorable if the following equivalent conditions hold:
\bea
|g|N & > & 6(\Omega-1) - \frac{(\Omega-1)^2}{\lambda} - 8\lambda, \nonumber\\
\lambda & < & \frac38(\Omega-1) -\frac{1}{16}|g|N -
\sqrt{\frac{1}{64}(\Omega-1)^2 -\frac{3}{64}|g|N(\Omega-1)
+\frac{1}{256}|g|N},\nonumber\\
\Omega & < & 1 + 3\lambda -\sqrt{\lambda(\lambda - |g|N)}.
\ena
If $\lambda -\sqrt{\lambda(\lambda - |g|N)} < |g|N/2$, then the
vortex state is favorable for all $\Omega > \Omega^v_{C1}$. If
$|g|N > \lambda$, then the ideal c.~m.\ state is the favorable one in the
whole interval $\Omega^v_{C1} < \Omega < \Omega^v_{C2}$ but the
vortex state may be favorable for higher frequencies, as discussed
above. To summarize the dependencies: (1) a small anharmonic term
$\lambda$ favors the ideal c.~m.-rotation state; (2) a small interaction term
$|g|N$ favors the vortex state; (3) a large angular velocity
$\Omega$ favors the vortex state.

\section{Equations of Motion}
\label{sec:equations}

The perturbative approach has, naturally, a limited range of
validity; as we shall see, it is in fact accurate only for very small
values of $gN$ and $\lambda$. There is thus need for a more
general scheme and we shall now discuss what kind of approximation
can be used to describe the attractive gas.

In a purely harmonic potential the center of mass motion decouples
from the internal motion. This is not the case in an anharmonic
trap, but there is still an approximate decoupling for weak
anharmonicity. By taking advantage of this, and the fact that the
internal motion is Bose condensed, coupled equations of motion for
the c.~m.\ and internal motion were derived in Ref.\ \cite{Us2003}:
\bea
   \left\{-\frac{\hbar^2}{2M}\frac{\partial^2}{\partial \RR^2}
   + V_C(\RR) \right\}\psi_C(\RR) = E_C \psi_C(\RR),
   \label{cme4}\\
   \left\{ -\frac{\hbar^2}{2m}
   \frac{\partial^2}{\partial \rr^2} + V_R(\rr)
   + U_0 |\Phi(\rr)|^2\right\} \Phi(\rr) = \mu \Phi(\rr).
\label{gpe4}
\ena
Here, the c.~m.\ is described quantum mechanically by the wave
function $\psi_C(\RR)$, and the Bose-Einstein condensed internal
motion is governed by the condensate wave function $\Phi(\rr)$,
where $\RR$ is the c.~m.\ coordinate and $\rr$ is the particle
coordinate relative to the c.~m. The effective potentials for the
center-of-mass and relative motion are
\bea
   V_C(R) = \frac12 M\omega^2(1+4\lambda \frac{\langle r^2\rangle}{\aos^2})
R^2 +
   \frac{\lambda M \omega^2}{2\aos^2} R^4,\label{cmpot}\\
   V_R(r) = \frac12 m\omega^2(1+4\lambda \frac{\langle R^2\rangle}{\aos^2}
   ) r^2 + \frac{\lambda m \omega^2}{2 \aos^2} r^4,
\label{gppot}
\ena
where $\langle R^2 \rangle = \int N^3 d\RR R^2|\psi_C(\RR)|^2$
and $\langle r^2 \rangle = N^{-1}\int d\rr r^2|\Phi(\rr)|^2$ gives
the coupling between the two wave functions. These equations are
capable of describing both the (ideal) state of c.~m.\ rotation and
vortex states and were used in Ref.\ \cite{Us2003} to find the
parameter regimes for those two states of motion. The
perturbative results of Sec.\ \ref{sec:perturbation} emerge from
these equations in the limit $\lambda\to 0, gN\to 0$ if $N$ is
assumed large.

Here, we shall take an alternative path to describing the motion
of the attractive gases. Namely, we argue that the ordinary
Gross-Pitaevskii equation \cite{Pethicksmith2001} can be used to
describe these rotating systems. The reason is simple: as shown
in Ref.~\cite{Pethick2000},
the state of c.~m.\ rotation is Bose-Einstein condensed
according to an observer co-moving with the c.~m. Therefore, if we
transform to a coordinate system moving with the c.~m., we should be
able to use the Gross-Pitaevskii equation. But if the cloud is in
its ground state, under a rotational force with frequency
$\Omega$, the c.~m.\ is also rotating with the frequency $\Omega$ and
the coordinate transformation is effected simply by going to a
rotating frame as usual \cite{Pethicksmith2001}:
\beq
H \to H' = H - \Omega \hat{L},
\enq
where $\hat{L}$ is the angular momentum operator. Solving the
Gross-Pitaevskii equation in the rotating frame should thus allow
us to describe stationary states of the cloud, both in the vortex
and the c.~m.\ states. This result comes as a bit of a surprise,
considering that in Ref.~\cite{Wilkin1998} the c.~m.\ state was found
{\em not} to be Bose-Einstein condensed. We note also that
working at fixed $\Omega$ is the correct description of the common
experimental approach of letting the cloud equilibrate under the
influence of a rotational drive.

Compared with the two coupled wave equations
(\ref{cme4}-\ref{gppot}), the full Gross-Pitaevskii equation is
more general: it lifts the restriction of small values of
$\lambda$, so that rotational motion in any trap can be described
as long as one can move to a reference frame in which the motion
is Bose-Einstein condensed. On the other hand, the GPE approach
describes the c.~m.\ motion classically instead of quantum
mechanically, but this is hardly an issue unless one wishes to
specifically study the quantum fluctuations in the c.~m.\ motion.
The quantum mechanical c.~m.\ motion will be illustrated in Sec.\
\ref{sec:correlations}.

\section{Numerical Simulations}
\label{sec:numerics}

As we argued in the preceeding section, the attractive Bose gas is
well described by the single component Gross-Pitaevskii
equation in a frame rotating with frequency ${\bf \Omega}$,
\beq
\label{GP}
\left[-\frac{\hbar^2}{2m}\nabla^2+
V\left(\rr\right)-
{\bf \Omega}\cdot\hat{L}+U_0|\Psi\left(\rr\right)|^2\right]
\Psi\left(\rr\right)
=\mu\Psi\left(\rr\right).
\enq
To solve this numerically for the anharmonic potential of
Eq.~(\ref{anharmpotential}), we choose $\lambda=0.15$, $N=1000$,
$\omega=2\pi\times 30$ Hz,
$\omega_z=2\pi\times 180$ Hz
and the mass of atomic Li$^{7}$.
For the harmonic trap frequency much stronger in $z$-direction,
we may assume the motion in this direction to be frozen in the ground
state of the trap and the problem becomes two-dimensional.
The ground state is now found by numerically propagating the
time-dependent counterpart of Eq.~(\ref{GP})
in imaginary time. The grid size varies
from $128\times 128$ points up to $256\times 256$.

\begin{figure}
\includegraphics[width=\columnwidth]{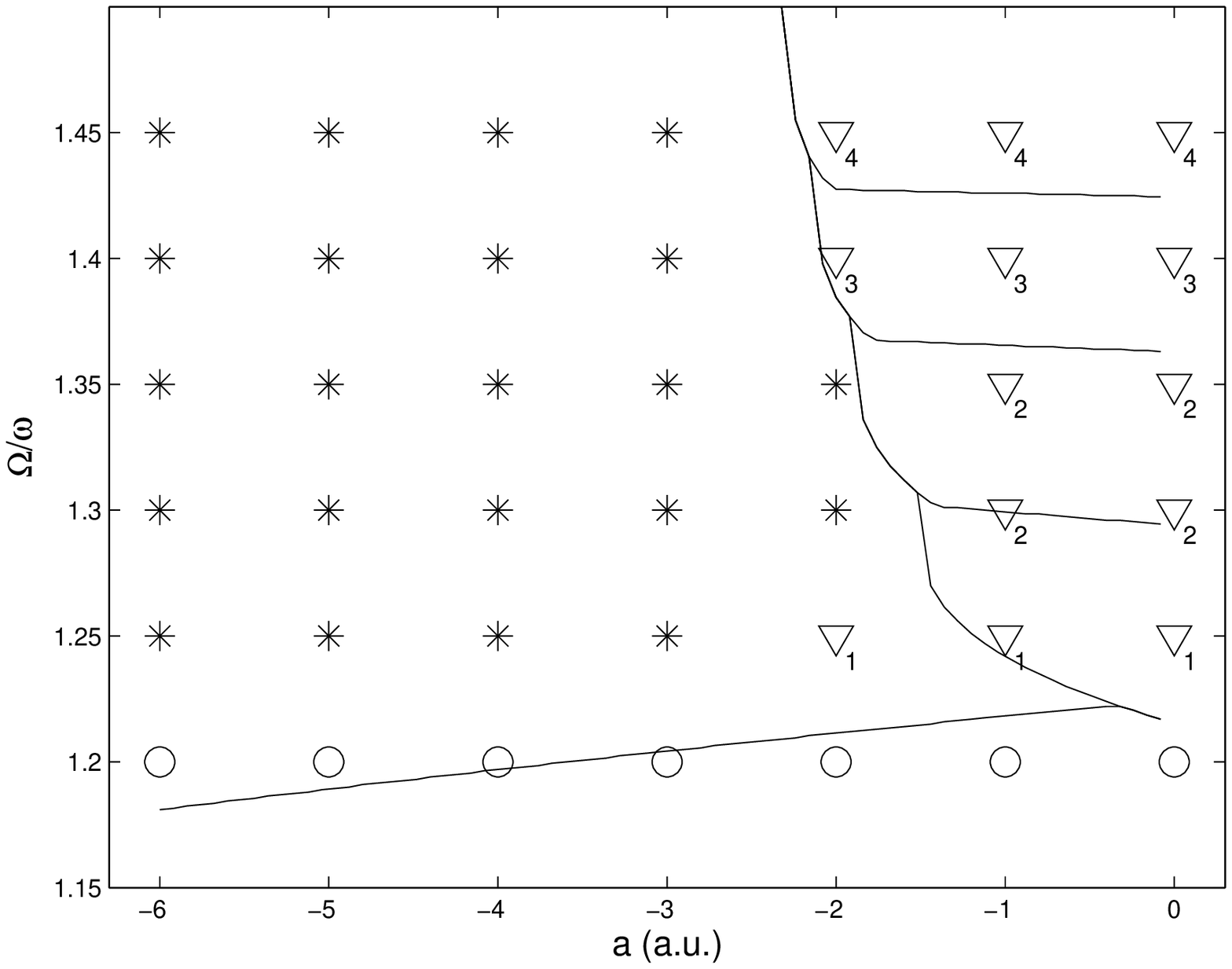}
\caption{The ground state phase-space diagram of a condensate in a rotating
trap with $\lambda=0.15$ as a function of the rotating frequency and the
scattering length. Different ground states are marked by the following
symbols: $\bigcirc$, nonrotating state; $\bigtriangledown$, vortex state
(the number is the circulation of a multiply quantized vortex); *,
c.~m.\ rotational state. The solid lines give the same phase-space diagram
obtained from the equations of motion where the c.~m.\ and internal motion are
decoupled, using a Gaussian ansatz for the wave functions.}

\label{pspace}
\end{figure}

With the fixed parameters mentioned above, we map the ground state
phase-space with respect to the rotational frequency and the
scattering length. The resulting phase-space diagram is plotted in
Fig.~\ref{pspace}. The result of the decoupling approximation,
Eqs.~(\ref{cme4}-\ref{gppot}), with a Gaussian ansatz for the
density \cite{Us2003} is also included. We see that the ground state is
nonrotating below 
$\Omega/\omega\approx 1.2$. The
cloud just stays at the bottom of the trap with angular momentum
$L = 0$ (within the numerical uncertainty). As the rotation
frequency is increased the ground state
depends on the interaction strength. For weak interactions 
the angular momentum $L/N$ is quantized in integer values that increase
with rotation frequency. The ground state is a (multiply)
quantized vortex (see Fig.~\ref{2ddens}).
On the contrary, stronger interactions break
the rotational symmetry. The circulation is not quantized anymore
and the density of the condensate has a shape of a crescent as can be seen
from Fig.~\ref{2ddens}. This configuration corresponds to the c.~m.\ rotational
state. For even stronger interactions, the density becomes more
concentrated and the cloud attains an approximately ellipsoidal
shape residing off center.

Interestingly, the coupled equations of motion
(\ref{cme4}-\ref{gppot}), derived from the factorization of the
wave function into c.~m.\ and internal parts, fail to describe the
elongated shapes that the cloud attains for moderately strong
attraction. The assumptions behind those equations imply, namely,
cylindrical symmetry of the condensate wave
function. The factorization ansatz thus fails precisely in the
situation when the quartic potential has a strong effect on the
shape of the cloud, i.e., when we are close to the phase boundary.

\begin{figure}
\caption{Density plots for different ground state configurations.
Bright shades indicate high density. On the left, we have a
doubly quantized vortex for $a=-1.0$ a.u., in the center
c.~m.\ rotational state for $a=-4.0$ a.u. and on the right c.~m.\ state for
$a=-10.0$ a.u. Anharmonicity is $\lambda=0.15$ and the rotation
frequency is fixed to $\Omega/\omega=1.35$. The unit of length for $x$
and $y$ is $\left(\hbar/m\omega\right)^{\frac{1}{2}}$.}
\label{2ddens}
\end{figure}

\begin{figure}
\includegraphics[width=\columnwidth]{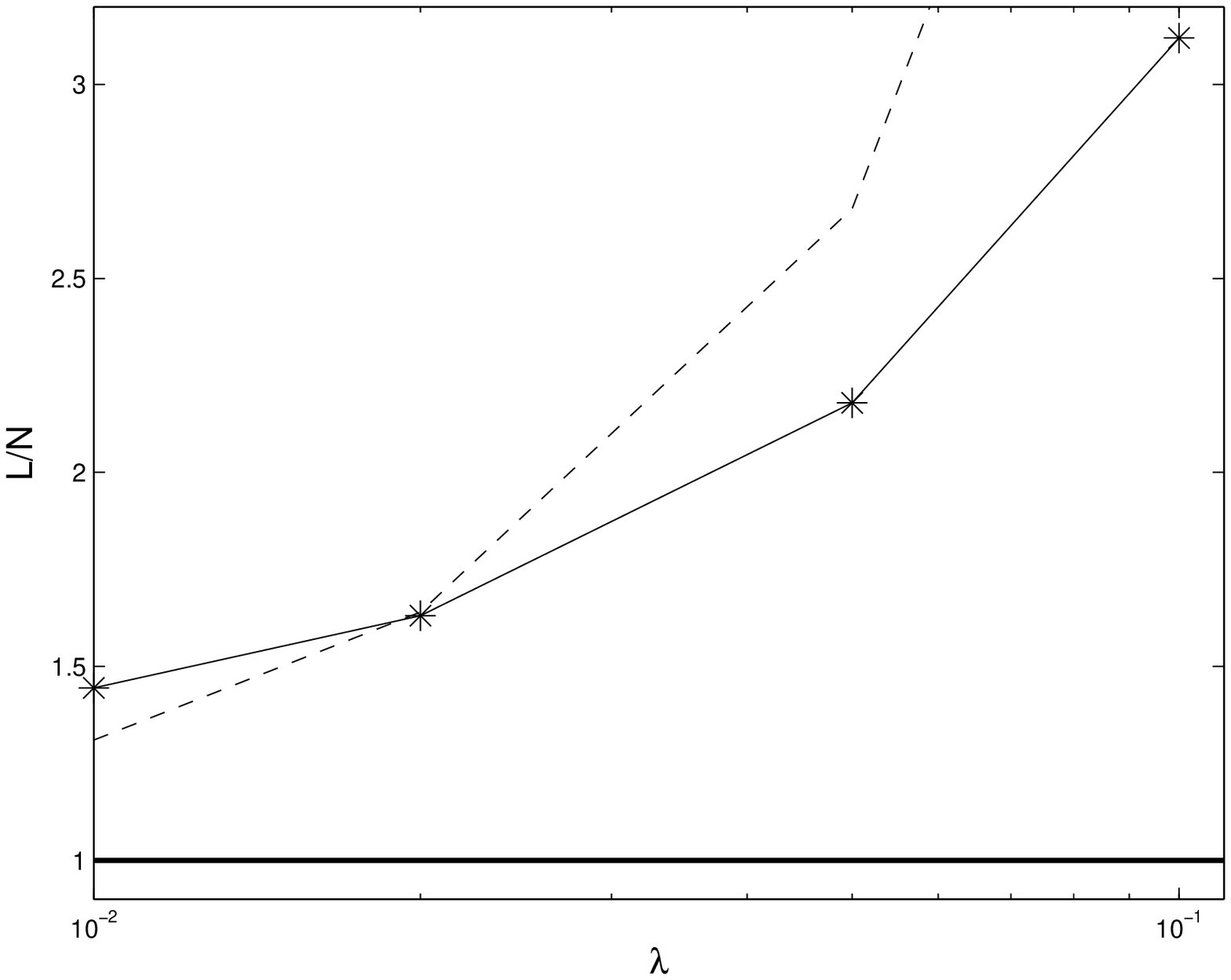}
\caption{The angular momentum (in units of $\hbar$) per atom as a 
function of the dimensionless
anharmonicity parameter $\lambda$ for the c.~m.\ state. 
The parameter $\Omega/\omega$ is chosen such that in 
perturbation theory $L/N=1$. Because we compare this to the results from
GP simulations (marked by the symbol *) we have to fix the dimensionless
interaction parameter $gN$ also. The 
parameter set for the lowest $\lambda$ is 
$\left(\lambda=0.01; \Omega=1.03; gN=0.015\right)$ and the others are 
$\left(0.02; 1.06; 0.03\right)$, $\left(0.05; 1.15; 0.075\right)$ and
$\left(0.1; 1.3; 0.15\right)$ respectively. The dashed line is variationally 
calculated for Gaussian trial wave functions in the decoupling approximation.}
\label{pertgp}
\end{figure}

From the perturbative method we have an expression for $q=L/N$ in
Eq.~(\ref{pertangmo}) as a function of $\Omega$ and $\lambda$.  We perform
a comparison between this simple perturbative formula for the ideal c.~m.\ 
state and the GP procedure by choosing the set of 
these pairs $\left(\Omega,\lambda\right)$ to 
be such that $q=1$. In Fig.~\ref{pertgp} we plot the results from the GP
simulations for the angular momentum per atom as a function of $\lambda$.
When the anharmonicity gets smaller $q$ seems to approach unity. Still, 
for the given parameter range the difference between numerical and perturbative
results is clear. This is somewhat unexpected, so to verify the 
numerical results we also plot the corresponding values of 
$q$ for Gaussian trial wave functions in the coupled 
Eqs. (\ref{cme4}-\ref{gppot}). The results are comparable 
to the ones from GP simulations; this shows that perturbation theory can only 
be trusted in an extremely small portion of phase space.

That the anharmonic trapping potential is essential for stabilizing these
states of motion can be seen from the following argument.
The effective potential seen by the rotating atoms is the sum of
the actual potential $V(r,\theta,z)$ and the centrifugal term:
\beq
V_{\rm eff}(r) = \frac12m(\omega^2-\Omega^2) r^2 +
\frac12m\omega^2\lambda\frac{r^4}{\aos^2}.
\enq
For slow rotation, $\Omega<\omega$, the prefactor of the $r^2$ term
is positive and the effective potential is
just a more shallow anharmonic potential. In that regime there is no
rotation for the attractive gas. For rotation faster than the trap frequency,
however, the effective potential attains a Mexican-hat
shape and the ground-state density distribution lies along the bottom of this
toroidal potential. If the trap bottom is deep and the effective trap width 
is small the condensate is confined in an effectively one-dimensional torus
which has been studied analytically in two recent
articles~\cite{Kana2003,Kavo2003}.
Just as in the present study, two types of state are found in
this idealized geometry, termed the uniform-density state and the
localized density state (bright soliton). Clearly, the uniform-density state
can be identified with our vortex state, and the bright soliton with
the c.~m.\ state. The conclusion drawn from the GP analysis is thus that 
the crossover from vortex to c.m.\ rotation is gradual, whereas 
the coupled equations (\ref{cme4}-\ref{gpe4}) predicted a discontinuous 
crossover \cite{Us2003}. In view of the arguments presented in 
Sec.\ \ref{sec:equations} to the effect that the GP equation provides a 
valid description of the gas, the correct conclusion is that the crossover 
is gradual. This conclusion is supported by the exact findings in the 
simplified 1D model \cite{Kana2003,Kavo2003}. The failure of the
coupled-equation approach can be ascribed to the fact that it is
restricted to cylindrically symmetric density profiles and is thus
unable to describe the elongated structures seen in Fig.\ \ref{2ddens}.

To check that the results are not specific to two dimensions, we have solved
the GP equation in 3D also. Both ground state configurations are obtained
in these simulations. As an example we present a density plot of a
c.~m.\ state in Fig.~\ref{3ddens}. Unfortunately, the 
computations are too time consuming for a quantitative 
mapping of the phase-space.

\begin{figure}
\caption{Density plot of a c.~m.\ rotational state obtained
by solving 3D GP equation. On the left $z=0$ plane and on the right
$y=0$ plane. The relevant parameters are $\Omega/\omega=1.40$, $a=-10.0$ a.u.
and $\lambda=0.15$. The unit of length for $x$, $y$ and $z$ 
is $\left(\hbar/m\omega\right)^{\frac{1}{2}}$.}
\label{3ddens}
\end{figure}

\section{Correlation analysis}
\label{sec:correlations}

It is instructive to visualize the dynamics of the many-body wave
function Eq.\ (\ref{wilkinwf}) found by Wilkin {\it et al.}
~\cite{Wilkin1998},
to see the connection between the wave function
and the result of the numerical Gross-Pitaevskii calculations
(Fig.\ \ref{2ddens}).
We therefore calculate the outcome of two consecutive measurements,
given by the correlation function
\beq
\label{corre}
c\left(z1,z2;t1,t2\right)=
\frac{\langle\hat{\psid}(z_1,t_1)\hat{\psid}(z_2,t_2)
\hat{\psi}(z_2,t_2)\hat{\psi}(z_1,t_1)\rangle}
{\langle\hat{\psid}(z_1,t_1)\hat{\psi}(z_1,t_1)\rangle},
\enq
that is, the conditional probability of detecting an atom at $(z_2,t_2)$
provided that one was detected at $(z_1,t_1)$. Here, we scale the lengths
and the energies as we did in Sec. \ref{sec:perturbation}. The time is
scaled by $1/\omega$.
By making a multi-nomial expansion, the many-body wave function
(\ref{wilkinwf}) can be expressed in the terms of
single-particle harmonic oscillator eigenstates:
\beq\label{wilkinwf2}
\Psi=\sum\sqrt{\frac{N^L}{\pi^N L!}}\frac{L!}{m_1!\cdots m_N!}
z_1^{m_1}\cdots z_N^{m_N}
\exp\left(\sum_{i=1}^N-|z_{i}|^2/2\right)
\enq
where the sum is taken over all possible combinations of $\{m_i\}$ that fulfill
$m_1+\cdots m_N=L$. This state can now be easily constructed
in the second quantized form
$|\Psi \rangle=\sum f\left(n_0,\ldots ,n_L\right) |n_0,\ldots ,n_L\rangle$
where the expansion coefficients are
\beq
f\left(n_0,\ldots ,n_L\right)=
\frac{1}{N^L}\left(\frac{N!}{n_0!\cdots n_L!}\right)^{\frac{1}{2}}
\left(\frac{L!}{\left( 0!\right)^{n_0}\cdots\left( L!\right)^{n_L}}\right)
^{\frac{1}{2}}
\enq
and $n_m$ denotes the population of atoms in the harmonic oscillator eigenstate
of angular momentum $m$.

\begin{figure}
\includegraphics[width=\columnwidth]{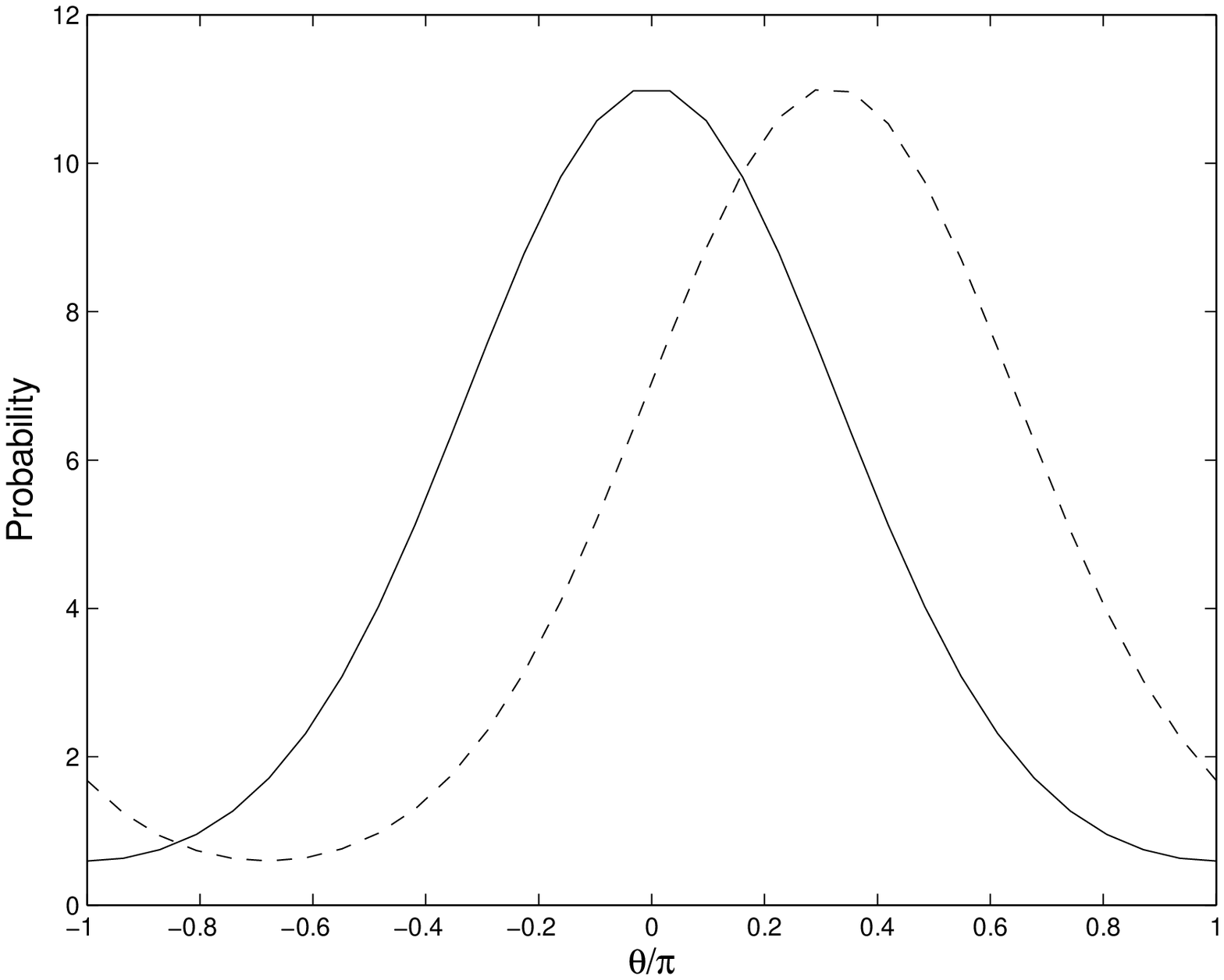}
\caption{Probability (unnormalized) of observing 
an atom at $\theta$ when another
atom was detected at $\theta=0$ for $t=0$. The solid line is the
time correlation Eq.\ (22) at $t=0$ and the dashed line represents
$t=1.0$. The unit of time is $1/\omega$.}
\label{fig:correplot}
\end{figure}

When the ideal c.~m.\ state is written as a superposition of 
different distributions of atoms in harmonic oscillator eigenstates, 
it is natural to determine the correlation (\ref{corre}) by 
writing the field operators in the harmonic oscillator basis
\beq
\hat{\psi}(z,t)=\sum_{m=0}^L\left(\pi m!\right)^{-\frac{1}{2}}
z^m\exp\left(-|z|^2/2\right)
\exp\left(-i(m+1)t\right)\hat{a}_m.
\enq
Here $\hat{a}_m$ is the bosonic particle destruction operator for
the state $m$. The form of the time dependent exponential comes from
the energy spectrum
\beq
E=\omega=|m|+2n_r+1.
\enq
In this case, to minimize the energy, the expansion
(\ref{wilkinwf2}) contains only terms where
the radial quantum number $n_r$ is zero.
We are interested in the dynamics as a function of the
polar angle $\theta$ with a fixed distance $R$ from the center of the trap.
Hence, we denote $z_1=Re^{i\theta_1}$ and $z_2=Re^{i\theta_2}$. In addition
we choose $t_1=0$, $\theta_1=0$, $t_2=t$ and $\theta_2=\theta$.
After doing some algebra, one gets
\beq
\langle\hat{\psid}(z_1,t_1)\hat{\psi}(z_1,t_1)\rangle=
\sum_f\sum_{m=0}^Le^{-R^2}
f^2\left(\pi m!\right)^{-1} R^{2m}n_m
\label{norm}
\enq
for the denominator. The numerator is a bit more tedious but
straightforward:
\bea
&&\langle\psid(z_1,t_1)\psid(z_2,t_2)\psi(z_2,t_2)\psi(z_1,t_1)\rangle=
\nonumber\\
&& \sum_f\sum_{m=0}^L\sum_{m'>m}^{L}\sum_{m''=0}^{L}
\frac{2f^2}{\pi^2}\frac{n_{m'} n_m}{m''!(m+m'-m'')!}
R^{2\left(m+m'\right)}e^{-2R^2}
\cos\left(\left(m-m''\right)\left(\theta-t\right)\right)+
\nonumber\\
&& \sum_f\sum_{m=0}^L\sum_{m'=0}^{m-1}
\frac{2f^2}{\pi^2}
\left[\frac{n_m\left(n_m-1\right)}{m'!\left(2m-m'\right)!}+
\frac{n_{m}n_{2m'-m}}{\left(m'!\right)^2}\right]
R^{4m}e^{-2R^2}
\cos\left(\left(m-m'\right)\left(\theta-t\right)\right)+
\nonumber\\
&& \sum_f\sum_{m=0}^L\sum_{m'>m}^L
\frac{2f^2n_{m'}n_m}{\pi^2m'!m!}
R^{2\left(m'+m\right)}e^{-2R^2}
\cos\left(\left(m'-m\right)\left(\theta-t\right)\right)+
\label{upst}\\
&& \sum_f\sum_{m=0}^L\sum_{m'>m}^L
\frac{2f^2n_{m'}n_m}{\pi^2m'!m!}
R^{2\left(m'+m\right)}e^{-2R^2}+\nonumber\\
&& \sum_f\sum_{m=0}^L
\frac{f^2n_m \left(n_m-1\right)}{\pi^2 \left(m!\right)^2}
R^{4m}e^{-2R^2}.\nonumber
\ena
In the first term $m''$ must satisfy certain conditions, namely
$m''\neq m$, $m''\neq m'$ and $m''\neq m+m'-m''$.
By combining (\ref{norm}) and (\ref{upst}), we get a slightly complicated
expression for the correlation function (\ref{corre}).
To visualize, we plot an example in
Fig.~\ref{fig:correplot} for $R=1$. 
The curves illustrate the conditional probabilities
at two different values of $t$ when $N=6$ and $L=6$. As
expected it describes a shape retaining a peaked structure moving 
clockwise at an angular frequency $\Omega=\omega$ (i.~e. $\Omega=1$
in dimensionless units).

\section{Conclusions}
\label{sec:conclusions}

In conclusion, we have studied the ground state properties of
an attractively interacting zero temperature Bose gas in the presence of
an anharmonic trap and a rotational drive. The two different ground state
configurations, vortex and center of mass rotation were investigated at
the limit when both anharmonicity and interactions are weak. In this limit,
analytical conditions for these states to appear were obtained as
functions of rotation frequency, anharmonicity and interaction strength.
To go beyond the perturbation theory, we have demonstrated that the ground
state can be solved by using the ordinary Gross-Pitaevskii
equation co-rotating with the external drive. 
This treatment differs from our previous
work~\cite{Us2003} where the coupled equations of motion were 
obtained by assuming the many body wave function to be a product of a c.m. 
wave function and a Bose condensed wave function of the motion relative 
to the c.m. Here, we treat the motion of the c.m. classically.
The Gross-Pitaevskii approach enables us to study any strength of the 
coupling or anharmonicity. We note that 
the comparison between different approximative methods shows that the 
perturbation result is valid for only a {\it very} limited parameter range.
The extension of the perturbative analysis performed in Ref. \cite{Kavo2004}
may improve the situation slightly.
Finally, we have visualized the c.m. dynamics by studying the correlations 
of two consecutive measurements.

The rotation of a Bose condensate with repulsive interactions in an
anharmonic trap is already experimentally performed \cite{Bretin2003}.   
The condensate was stirred with a laser in a trap which was 
created by combining a harmonic magnetic potential with 
an optical potential of Gaussian shape. In principle, by using a 
similar technique for attractively interacting 
Bose gas one could be able to observe the rotational ground states we
have studied. The possibility to tune scattering 
lengths with magnetic fields via Feshbach resonances 
\cite{Feshb1992,Tiesi1993} offers another tool for probing 
the parameter space.

\section{Acknowledgments}
Discussions with C.~J.\ Pethick and G.\ Kavoulakis are
gratefully acknowledged. We also thank J.-P. Martikainen for valuable
help in the numerical procedure and M. Mackie for critical reading
of the manuscript. The authors acknowledge support from the
Academy of Finland (Project 206108), the European Network
``Cold Atoms and Ultra-Precise Atomic Clocks'' (CAUAC), and
the Magnus Ehrnrooth foundation.

\end{document}